\documentstyle[11pt,aaspp4]{article}

\input psfig
\psfull
\singlespace

\parskip=\the\medskipamount

\def\etal{{\it et al. \/}}

\def\Nhosts    {N_{H}({\mathcal{M}})}
\def\Ntotal    {N({\mathcal{M}})}
\def\PZdestroy {P_{DE}({\mathcal{M}})}
\def\PZform    {P_{PE}({\mathcal{M}})}
\def\PZsurvive {P_{HE}({\mathcal{M}})}

\newcommand {\lsim}{\mbox{$\:\stackrel{<}{_{\sim}}\:$} }
\def\be{\begin{equation}}
\def\ee{\end{equation}}
\def\bea{\begin{eqnarray}}
\def\eea{\end{eqnarray}}

\begin{document}
\title{
An Estimate of the Age Distribution of Terrestrial Planets\\
in the Universe: Quantifying Metallicity as a Selection Effect}
\medskip 

\author{Charles H. Lineweaver\\
School of Physics, University of New South Wales\\
charley@bat.phys.unsw.edu.au\\
Icarus, in press}

\begin{abstract}

Planets like the Earth   
cannot form unless elements heavier than helium 
are available. These heavy elements, or `metals', were not produced in the big bang. 
They result from fusion inside stars and have been gradually building up
over the lifetime of the Universe.
Recent observations indicate that the presence of giant extrasolar planets
at small distances from their host stars, is strongly correlated with 
high metallicity of the host stars.
The presence of these close-orbiting
giants is incompatible with the existence
of earth-like planets.
Thus, there may be a Goldilocks selection effect:
with too little metallicity, earths are unable to form for lack of material, with too 
much metallicity giant planets destroy earths.
Here I quantify these effects and obtain the probability, as a function of metallicity, for
a stellar system to harbour an earth-like planet.
I combine this probability with current estimates of the star formation rate 
and of the gradual build up of metals in the Universe to obtain an estimate of the
age distribution of earth-like planets in the Universe.
The analysis done here indicates that
three quarters of the earth-like planets in the Universe are older than the Earth and that their
average age is $1.8 \pm 0.9$  billion years older than the Earth.
If life forms readily on earth-like planets 
-- as suggested 
by the rapid appearance of life on Earth -- this analysis gives us an age 
distribution for life on such planets and a rare clue about how we 
compare to other life which may inhabit the Universe. 

\end{abstract}



\section{Aims}

Observations of protoplanetary disks around young stars in star-forming regions
support the widely accepted idea that planet formation is a common by-product 
of star formation (e.g. Beckwith  \etal 2000).
Our Solar System may be a typical planetary system in which
earth-like planets accrete near the host star from rocky debris depleted of volatile elements, while 
giant gaseous planets accrete in the ice zones ($\gtrsim 4$ AU) around rocky cores (Boss 1995, Lissauer 1996).
When the rocky cores in the ice zones reach a critical mass ($\sim 10 \: m_{Earth}$) runaway gaseous accretion
(formation of jupiters) begins
and continues until a gap in the protoplanetary disk forms or the disk dissipates 
(Papaloizou and Terquem 1999, Habing \etal 1999). 
The presence of metals is then a requirement for the formation of both earths and jupiters.

We cannot yet verify if our Solar System is a typical planetary system or how generic the pattern described above
is. 
The Doppler technique responsible for almost all extrasolar planet detections
(Mayor \etal 1995, Butler \etal 2000 and references therein)
is most sensitive to massive close-orbiting planets and is
only now becoming able to detect planetary systems like ours, i.e., jupiters at $\gtrsim 4$ AU from nearby host stars.
The Doppler technique has found
more than 40 massive ($0.2 \lsim m/m_{Jup} \lsim 10$) extrasolar planets in close 
($0.05 \lsim a \lsim 3$  AU), often  eccentric orbits around 
high metallicity host stars (Schneider 2000).
I refer to {\it all} of these giants as `hot jupiters' because of their high mass and proximity to their central stars.
Approximately 5\% of the sun-like stars surveyed 
possess such giant planets (Marcy and Butler 2000).
Thus there is room in the remaining 95\% for stars to harbour planetary systems like
our Solar System.

It is not likely that giant planets have formed in situ so close to their host stars (Bodenheimer \etal 2000).
It is more likely that after formation in the ice zone, these giants moved through the habitable zone, destroyed nascent earths
(or precluded their formation) and are now found close to their host stars (Lin \etal 1996).
How this migration occurred is an active field of research.  
However, independent of the details of this migration, recent detections of extrasolar planets are telling us more about 
where earths are not, than about where earths are.

The aims of this paper are to use the most recent observational data to quantify
the metallicity range compatible with the presence of earths 
and estimate 
the age distribution of earth-like planets in the Universe.
The outline of the analysis is as follows:

\begin{enumerate}
\item
compare the metallicity distribution of stars hosting hot jupiters with
the metallicity distribution of stars in the solar neighborhood to obtain the probability of hosting hot jupiters 
(and therefore the probability of destroying earths)
\item
assume that 
starting in extremely low metallicity stars, the probability to produce earths increases linearly with metallicity 
(this assumption is discussed in Section 2.2)
\item
combine items 1 and 2 above to estimate the probability of harbouring earths as a function
of metallicity (Fig. 1)
\item
use current estimates of the star formation rate in the Universe  (Fig. 2A) and observations of high redshift metallicities 
to estimate the metallicity distribution of star-forming regions as a function of time (Fig. 2B)
\item
combine 3 and 4 above to estimate the age distribution of earth-like planets in the Universe (Fig. 2C)
\end{enumerate}

\begin{figure*}[!ht]
\centerline{\psfig{figure=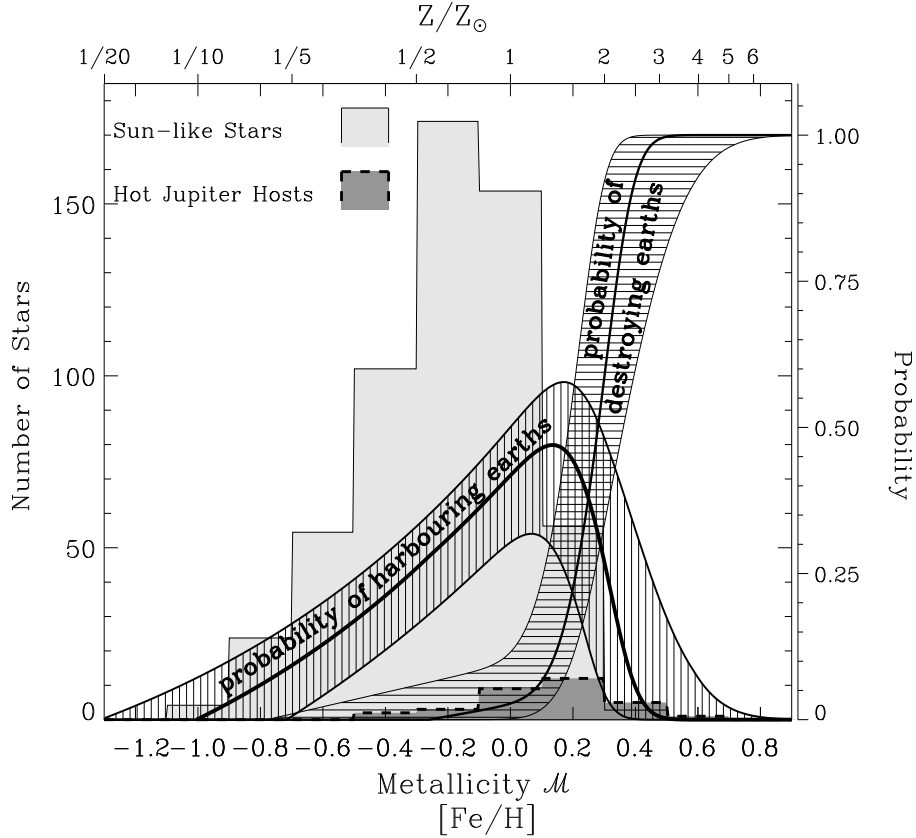,height=13.0cm,width=12.5cm}}
\caption{
If metallicity had no effect on planet formation we would expect the metallicity distribution
of stars hosting hot jupiters (giant, close-orbiting, extrasolar planets, dark grey)
to be an unbiased subsample of the distribution of sun-like stars in the solar neighborhood (light grey). 
However, hot jupiter hosts are more metal-rich.
Hot jupiters have the virtue of being Doppler-detectable but they destroy or preclude
the existence of earths in the same stellar system.
For a given metallicity, the probability of destroying earths is the ratio of the number of 
hot jupiter hosts to the number of stars surveyed (Eq. 1).
The probability of harbouring earths (Eq. 2) is
based on the assumption that the production of earths is
linearly proportional to metallicity, but is cut off at high metallicity by the 
increasing probability to destroy earths.
The upper x-axis shows the linear metal abundance.
 }
\label{fig:metals}
\end{figure*}
\section{Harbouring and Destroying Earths}

Figure 1 shows the metallicity distribution of 32 
stars hosting hot jupiters whose metallicities 
have been published  (Gonzalez 2000, Table 1, Butler \etal 2000, Table 4 and references therein).
The fact that these hosts are significantly more metal-rich than sun-like stars in the solar neighborhood
has been reported and discussed in several papers, including Gonzalez (1997, 1998, 2000), 
Ford, Rasio and Sills (1999), Queloz \etal (2000), Butler \etal (2000).
The metallicity distribution of sun-like stars in Fig. 1 is a linear combination of
similar histograms in  Sommer-Larsen (1991) and Rocha-Pinto and Maciel (1996).
These two references were chosen because their G dwarf samples are taken from the solar 
neighborhood and, although they are not identical to the metallicity distribution 
of the target stars that have been searched for planets using the Doppler technique,
they are good representatives of these stars.

Let the observed metallicity distribution of the sun-like stars hosting 
giant planets be $\Nhosts$, where ${\mathcal{M}}= [Fe/H] \equiv log(Fe/H) - log(Fe/H)_{\odot} \approx log(Z/Z_{\odot})$,
and $Z_{\odot} = 0.016$ is the mass fraction of metals in the Sun.
In Fig. 1, the distribution of sun-like stars, $\Ntotal$, 
has been normalized so that the 32 host stars represent
$5.6\%$ (the average planet-finding efficiency given in Marcy and Butler 2000) of the total.
That is, each bin of $\Ntotal$ has been rescaled such that 
$0.056 \: \sum_{i} N({\mathcal{M}}_{i})= 32$.
Target stars have not been selected for metallicity. Although Doppler shifts can be measured with slightly more
precision in metal-rich stars, and metal-rich stars are slightly brighter for a given spectral type 
(leading to a Malmquist bias), these two selection effects are estimated to be minor compared to the 
difference between the distributions in Fig. 1 (Butler \etal 2000).  

\subsection{Probability of destroying earths}

For a given metallicity, an estimate of the relative probability that a star will host a hot jupiter 
is the ratio of the number of stars hosting hot jupiters to the number of stars targeted,

\be
\PZdestroy  = \frac{\Nhosts}{\Ntotal}.
\ee

This is plotted in Fig. 1 and labelled ``probability of destroying earths''.
At low metallicity, $\PZdestroy$ is low and remains low until solar metallicity, where it rises steeply.
This probability predicts that more than $95\%$ of sun-like stars with ${\mathcal{M}}> 0.4$ will have a 
Doppler-detectable hot jupiter, $\sim 20\%$ of ${\mathcal{M}}\sim 0.2$ stars will have one and
$\sim 5\%$ of solar metallicity stars (${\mathcal{M}} \sim {\mathcal{M}}_{\odot} \equiv  0$) will have one.
These predictions are also supported by independent observations:
\begin{itemize}
\item
A star with extremely high metallicity was included in the target list because of its high
metallicity (${\mathcal{M}} = 0.5$). A planet, BD-10 3166, was found around it (Butler \etal 2000).
This star was not included in Fig. 1 because of selection bias, but this result does support
the probability calculated here: $P_{DE}({\mathcal{M}} = 0.5) \sim 1$. 

\item
Thirty-four thousand stars in the globular cluster Tucanae 47 (${\mathcal{M}}= -0.7$) were monitored with HST for
planets transiting the disks of the hosts.
Fifteen or twenty such transits were predicted based on a $\sim 5\%$ planet-finding efficiency 
(assumed to be independent of metallicity and stellar environment).
None has been found (Gilliland \etal 2000). This result is consistent with the 
probability calculated here: $P_{DE}({\mathcal{M}}=-0.7) \sim 0$, but Gilliland \etal (2000)
also suggest that the lack of planets could be due to high stellar densities disrupting planetary stability.

\end{itemize}

The width of the `probability of destroying earths' region has been set by the errors on the
terms in Eq. 1.
The region is broad enough to contain $\PZdestroy$'s calculated when 
alternative estimates of $\Ntotal$ are used singly or in combination 
(e.g. Sommer-Larsen 1991, Rocha-Pinto and Maciel 1996, Favata \etal 1997) and 
when a range of planet-finding efficiencies are assumed ($3\%$ to $10\%$).
Thus, the curve is fairly robust to variations in both the estimates of the metallicity distribution
of the target stars and to varying estimates of the efficiency of finding hot jupiters.
When larger numbers of hot jupiters are found and the metallicity distribution of the target stars
is better known, the new $\PZdestroy$ should remain in (or very close to) the region labelled
``probability of destroying earths'' in Fig. 1.

\subsection{The probability of producing earths}

The probability of producing earths is zero at zero metallicity and increases as metals build up in the Universe. 
The qualitative validity of this idea is broadly agreed upon 
(Trimble 1997, Whittet 1997) but it is difficult to quantify.
During star formation, varying degrees of fractionation transform a stellar metallicity disk
into rocky and gaseous planets.
Simulations of terrestrial planet formation by Wetherill (1996) suggest that the mass of rocky 
planets within 3 AU is approximately proportional 
to the surface density of solid bodies in the protoplanetary disk.
In the low surface density regime, ($\sim 3 \; g/cm^{2}$), where finding enough material 
to make an earth is a problem,
the number of planets in the mass range $\;0.5 < m/m_{earth} < 2$ increases 
roughly in proportion to the surface density, i.e., to the metallicity. This increase is not 
because the overall number of planets increases 
but because the masses (of a constant number of planets) increase, bringing them into the earth-like
mass range.
These simulations may be the best evidence we currently have to support the idea that 
in the low metallicity regime,
the probability of
forming earths is linearly proportional to metallicity.
This also suggests that the earliest forming earths orbit minimal metallicity stars 
and are at the low mass end of whatever definition of `earth-like' is being used.

Similar considerations apply to jupiter formation
but at a slightly higher surface density threshold.
Weidenschilling (1998) finds that a $10 \; g/cm^{2}$ disk surface density is not quite enough to initiate 
runaway jupiter formation but that a modest increase in surface density will.
These simulations support the standard core accretion models of planet formation and suggest that 
planet formation (both rocky and gaseous) is 
enhanced when more metals are available.

In this analysis I make the simple assumption that
the ability to produce earths is zero at low metallicity
and increases linearly with metallicity of the host star.  
Specifically, let $\PZform$  be the relative ability to produce earths as a function of metallicity.
I assume that:
\begin{itemize}
\item
$\PZform \propto Z$ (earth production is proportional to the abundance of metals) 
\item
$P_{PE}({\mathcal{M}}= -1.0) = 0$.
That is, at very low metal abundance, $(Z/Z_{\odot} \sim 1/10)$, the probability of producing earths is $0$. 
To represent the uncertainty in this zero probability boundary condition, the range $1/20 < Z/Z_{\odot} < 1/5$ 
is shown in Fig. 1. This assumption is discussed later.

\item The most metallic bin,  ${\mathcal{M}} = 0.6$, is assigned the probability of 1: $P_{PE}({\mathcal{M}}=0.6) = 1$.

\end{itemize}

\subsection{Probability of harbouring earths}

The probability of a stellar system harbouring earths, $\PZsurvive$, is 
the probability of producing earths times the probability of not destroying them,

\be
\PZsurvive = \PZform \:* [1- \PZdestroy].
\ee

This probability of harbouring earths is plotted in Fig. 1.
Starting at low metallicity, it rises linearly and then gets cut off sharply at ${\mathcal{M}} \gtrsim 0.3$.
It peaks at ${\mathcal{M}} = 0.135$,
has a mean of $ -0.063$ and a median of $-0.036$.
The $68\%$ confidence range is $[-0.38 < {\mathcal{M}} < 0.21]$. 
%
$\PZsurvive$ can be used to focus terrestrial planet search strategies.
For example, to maximize the chances of finding earths, NASA's terrestrial planet finder (TPF) should
look at stars with metallicities within the 68\% confidence range and in particular
near the peak of $\PZsurvive$.
Also, since there is a radial metallicity gradient in our galaxy, $\PZsurvive$ can be used to
define a galactic metallicity-dependent habitable zone analogous to the water-dependent
habitable zones around stars. This can be done by replacing `$t$' in Eqs. 3, 4 and 6 with galactic radius
`$R$' and replacing the $SFR(t)$ with the density of sun-like stars $\rho(R)$.

The Sun  (${\mathcal{M}}_{\odot} \equiv [Fe/H] \equiv 0$) is more 
metal-rich than $\sim 2/3$ of local sun-like stars and less metal-rich than $\sim 2/3$ of
the stars hosting close-orbiting  extrasolar planets.
The high value of ${\mathcal{M}}_{\odot}$ (compared to neighboring stars) 
and the low value compared to hot jupiter hosts
may be a natural consequence of the Goldilocks metallicity selection effect discussed here
(see also Gonzalez 1999).

Models need to simultaneously explain
the presence of hot jupiters close to the host star, the high metallicity of the host stars 
(specifically the steepness of the rise in
$P_{DE}$ seen in Fig. 1)
as well as the small eccentricities of the closest orbiting planets and the large eccentricities of the 
planets further out.
Planet/planet gravitational scattering may provide a natural way to explain these features
(Weidenschillling and Marzari 1996, Rasio and Ford 1996 and Lin and Ida 1997).
Higher metallicity of the protoplanetary disk enhances the mass and/or
number of giant planets, thereby enhancing the frequency of gravitational encounters between them.
Simulations with up to nine planets have been done (Lin and Ida 1997) and apparently, ``the more the merrier''.
That is, the more planets there are, the more likely one is to get scattered into a sub-AU orbit.
The least massive planets suffer the largest orbital changes. Thus the least massive
are more likely to be ejected, but also are more likely to be gravitationally
scattered into orbits with small periastrons which can become partially circularised either
by tidal circularization or by the influence of a disk (provided the disk has not dissipated before
the scattering).

\section{Star Formation Rate of the Universe}

The observational determination of the star formation rate ($SFR$) of the Universe has been the focus of much 
current work which we summarize in Fig. 2 A.  
Various sources indicate a $SFR$ at high redshift an order of magnitude larger than the current
$SFR$.
Initial estimates in which a peak of star formation was found at redshift $1 \lsim z \lsim 2$
are being revised as new evidence indicates that there may be no peak in the
star formation rate out to the maximum redshifts available ($z \sim 5$).

Let us restrict our attention to the set of sun-like  stars
(spectral types F7 - K1, in the mass range $0.8 \lsim m/m_{\odot} \lsim 1.2$) that
have ever been born in the Universe.
Since $\sim 5\%$ of the mass that forms stars forms sun-like stars,
the star formation rate as a function of time, $SFR(t)$ can be multiplied by $A \sim 0.05$ to yield 
the age distribution of sun-like stars in the Universe.
Here, the standard simplification is made that the stellar initial mass function is constant.
If low mass star formation is suppressed in low metallicity molecular clouds (Nishi and  Tashiro 2000)
then the 1.5 Gyr delay between $SFR$ and $PFR$  (Section 5) is even longer.

If all sun-like stars formed planets irrespective of their metallicity, then the planet formation

\clearpage
\vspace{-2cm}
\begin{figure*}[!ht]
\centerline{\psfig{figure=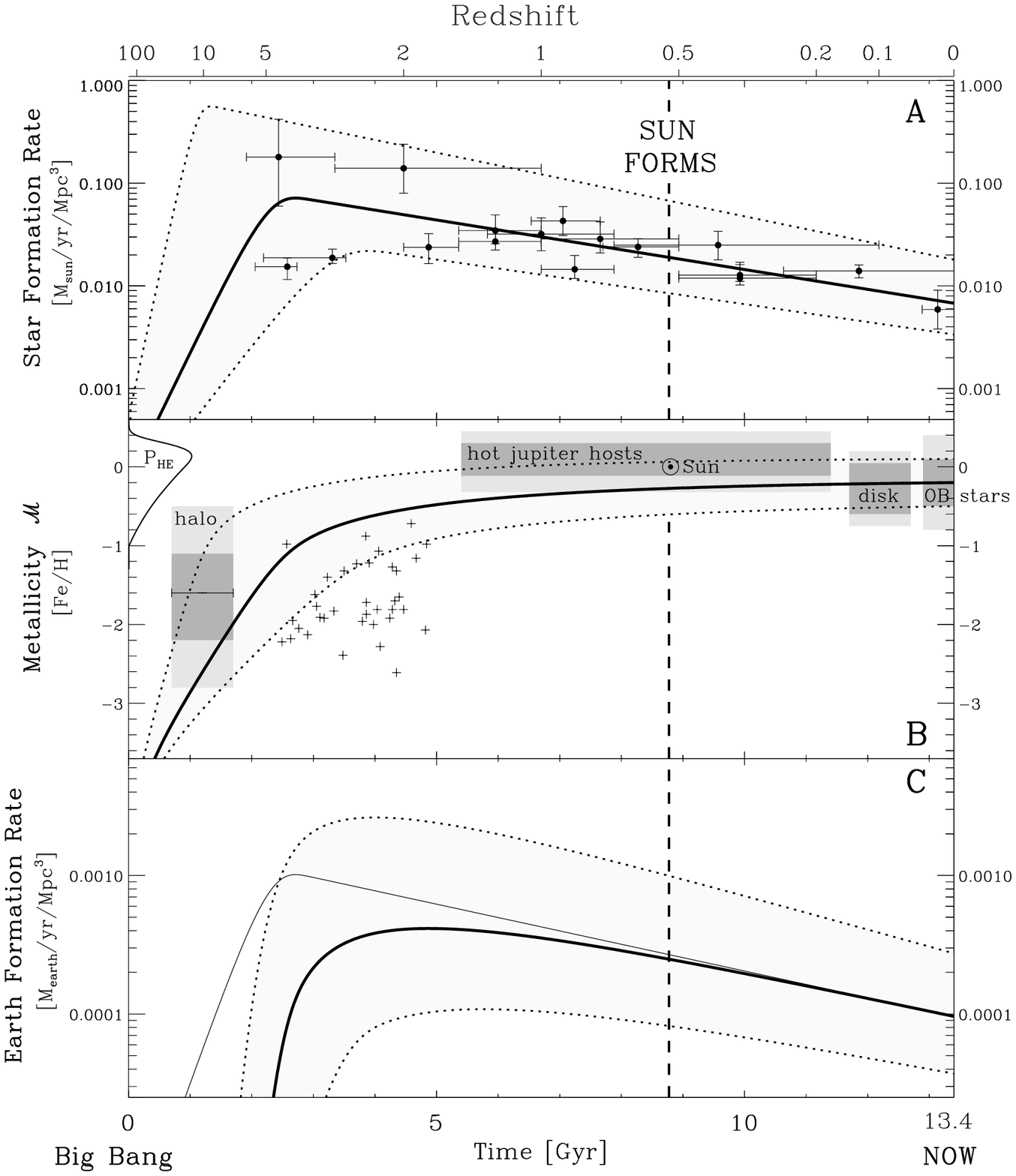,height=15.0cm,width=16.cm}}
\vspace{-1cm}
\caption{
{\footnotesize
These three panels show ({\bf A}) current estimates for the evolution of the star formation rate in the Universe,
({\bf B}) current estimates of the build up of metallicity in the Universe and
({\bf C}) the age distribution of earths in the Universe.
The cumulative effect of star formation is to gradually increase the metallicity of the Universe.
The cumulative integral of the star formation rate in {\bf A} (Eq. 5) is plotted in {\bf B}.
A combination (Eq. 4) of  the probability of harbouring earths (Eq. 2) 
with the metallicity of the Universe as a function of time (Eq. 6)
yields an estimate (Eq. 3) of the age distribution of earths in the Universe ({\bf C}).
The star formation rates in {\bf A} are from a compilation in Barger \etal (2000).
The grey band represents the uncertainty in the star formation rate and controls the width of the
grey bands in {\bf B} and {\bf C}.
In {\bf B}, the metallicity distributions of
various stellar populations are plotted and are consistent
with this universal metallicity plot.
The metallicity distribution of the stars in the Milky Way halo (Laird \etal
1988) is represented by the dark grey (68\% confidence level) and light grey
(95\% confidence level) and is plotted at its time of formation (Lineweaver 1999).
The metallicity distributions of stars in the
Milky Way disk (Favata {\it et al.\/} 1997),
of massive OB stars (Gummersbach \etal 1998) and of stars hosting hot jupiters
are similarly represented. 
The probability of harbouring earths, $P_{HE}$ from Fig. 1 is plotted
in the top left of {\bf  B}.
The ``+'' signs in {\bf  B} are the metallicity of damped Lyman-$\alpha$ 
systems from a compilation by Wasserburg and Qian (2000).
The age range of the disk metallicity has been reduced to aid comparison with the 
OB stars and the hot jupiter hosts.
The thin solid line in {\bf C} is the star formation rate from {\bf A}, rescaled to the current earth formation
rate.
If the formation of earths had no metallicity dependence (or any other dependence on a time-dependent quantity)
it would be  identical to such a rescaling of the star formation rate.
The $\sim 1.5$ Gyr delay between the onset of star formation and the onset of earth formation
is due to the metallicity requirements for earth formation.}
}
\label{fig:sfr}
\end{figure*}
%
%
\noindent rate in the Universe would equal $A\: * \: SFR(t)$, shown in Fig. 2A.
However, Fig. 1 indicates that metallicity is a factor which should be taken into account.
Thus, we estimate the earth-like planet formation rate, $PFR(t)$, orbiting sun-like stars as 

\be
PFR(t) = A \:*\: SFR(t) \;*\: f(t),
\ee

\noindent where $f(t)$  is the fraction of stars being formed at time $t$ which are able to harbour earths.
If all sun-like stars formed planets and metallicity had no effect on terrestrial planet formation,  
we would have $f(t) = 1$.
If we knew that on average one out of every thousand sun-like  stars 
had an earth-like planet, (and this number did not depend on the metallicity 
of the star), then the planet formation rate would be $PFR(t) = A \: * \:  SFR(t) \: * \: 0.001$, 
which is just a rescaling of the $SFR$ in Fig. 2A.
A plausible first approximation could have
$f(t) \propto \overline{{\mathcal{M}}}(t)$.
That is, the higher the average metallicity of the Universe, the higher the efficiency with which
star formation produces earths.
In this analysis, however, this guess is improved on by taking into account the dispersion of metallicity
of star forming regions around the mean at any given time, as well as by including the 
metallicity dependent selection effect for harbouring earths.
When these are taken into consideration, $f(t)$ becomes an integral over metallicity,

\be
f(t) \approx \int \; P({\mathcal{M}}, \overline{{\mathcal{M}}}(t)) \; \PZsurvive\; d{\mathcal{M}},
\ee

\noindent where $P_{HE}$ was derived above and 
$P({\mathcal{M}},\overline{{\mathcal{M}}}(t))$ is a Gaussian parametrisation of the 
metallicity distribution of star-forming regions in the 
Universe (Eq. 6).

\section{The Build-up of Metallicity in the Universe}

The Universe started off with zero metallicity and a complete inability to
form earths. The metallicity of the Universe gradually increased as
a result of star formation  and its by-products: various types of stellar novae and stellar winds.
%
Various observations form a  consistent picture of the gradual increasing metallicity of the Universe
(Fig. 2B).

The star formation rate plays a dual role in this analysis since stars make planets ($PFR \propto SFR$, Eq. 3)
and stars make metals, $\frac{d{\mathcal{M}}}{dt}(t) \propto SFR(t)$. Integration of this last proportionality 
yields the increasing mean metallicity, $\overline{{\mathcal{M}}}(t)$, of star forming regions in the Universe
\be
\int_{0}^{t} SFR(t')dt' \sim \overline{{\mathcal{M}}}(t).
\ee

\noindent The resulting $\overline{{\mathcal{M}}}(t)$, is plotted in Fig. 2B (thick line).
%
The grey area around $\overline{{\mathcal{M}}}(t)$ reflects the spread in the estimates of the $SFR$ (grey area in ${\bf A}$).
Metallicity observations in our galaxy and at large redshifts are available to check the plausibility of this integral
and are shown in Fig. 2B.

At any given time $t$, some star forming regions have low metallicity while some have high metallicity. We parametrize this
spatial dispersion around the mean by a time-dependent Gaussian centered on $\overline{{\mathcal{M}}}(t)$: 

\be
P({\mathcal{M}},\overline{{\mathcal{M}}}(t)) = \frac{1}{\sigma \sqrt{2\pi}}
exp{[\frac{({\mathcal{M}} - \overline{{\mathcal{M}}}(t))^{2}}{2\sigma^{2}}]}.
\ee

\noindent The current metallicity distribution of OB stars in the thin disk (Gummersbach \etal 1998),
which may be our best estimate of the current mean 
metallicity of star-forming regions in the Universe,   
is used to normalize this function, $\overline{{\mathcal{M}}}(t_{o}) = 0.63$, and provide the dispersion, 
$\sigma = 0.3$.

\section{The Age Distribution of Earths in the Universe}

Performing the integral in Eq. 4 and inserting the result into Eq. 3 yields
an estimate of the terrestrial planet formation rate in the Universe which is also
the age distribution of earths orbiting sun-like stars in the Universe.
This distribution is plotted in Fig. 2C and indicates that
the average age of earths around sun-like stars is $6.4 \pm 0.9$ 
billion years.
The error bar represents the uncertainty in the $SFR$ (shown in Fig. 2A)
as well as the range of assumptions about the low metallicity tail of $P_{FE}$, discussed below.
Thus, the average earth in the Universe is $1.8 \pm 0.9$
billion years older than our Earth.
And, if life exists on some of these earths, it will have evolved, on average, $1.8$ billion years
longer than we have on Earth.
Among these earths, $74 \pm 9 \%$ are older than our Earth while $26 \pm 9 \%$ are younger.
%
$68\%$ of earths in the Universe are  between $3.3$ and $9.3$ Gyr old while $95\%$ are 
between $0.6$ and $10.5$ Gyr old. 


The time delay between the onset of star formation and the onset of terrestrial planet formation is 
the difference between the x-intercept of the thin and thick solid lines in Fig. 2C.
This delay is $\sim 1.5 \pm 0.3$ Gyr 
and has an important dependence on the low metallicity tail of $P_{HE}$, specifically, on the metallicity 
for which $P_{FE}({\mathcal{M}}) = 0$ has been assumed.
To estimate the dependence of the main result on this assumption, both a high and
a low metallicity case ($Z/Z_{\odot} = 1/5 $ and $1/20$) have been considered. 
That is, I have used the two boundary conditions $P_{FE}({\mathcal{M}} = -0.7) = 0$
and $P_{FE}({\mathcal{M}} = -1.3) = 0$ and the variation they produce in the result,
to compute representative error bars.
The resulting variation is about one half of the variation due to
the uncertainty in the $SFR$.
If rocky planets can easily form around stars with extremely low metallicity due to high levels of
fractionation during planet formation, then the lower limit used here, $Z/Z_{\odot} =1/20$, may not be low enough.
%

These linear metallicity variations yield error estimates but
other possibilities exist.
The masses of earth-like planets and the ability of a stellar system to produce them
may not be linear functions of metallicity. 
For example, there may be a strongly non-linear dependence on metallicity such 
as a metallicity threshold below which earths do not form and above which they always do.
If that were the case then the $PFR$ plotted in Fig. 2C
would shift to the right or left depending on where the threshold is.

In this analysis I have assumed that the moons of hot jupiters do not accrete into earth-like
planets. This speculation has not been explored in any detail. If true, 
hot jupiters would destroy earths, but would also help create alternative sites for life.
However, the delayed onset of planet formation compared to star formation derived here
would be largely unchanged.

The cratering history of the Moon tells us that the Earth underwent an early intense
bombardment by planetesimals and comets from its formation 4.56 Gyr ago until
$\sim 3.8$ Gyr ago.
For the first 0.5 Gyr, the bombardment was so intense (temperatures so high) that
the formation of early life may have been frustrated  (Maher and Stevensen 1988).
The earliest isotopic evidence for life dates from the end of this heavy bombardment
$\sim 3.9$ billion years ago (Mojzsis \etal 1996). Thus, life on Earth seems to have arisen
as soon as temperatures permitted (Lazcano and Miller 1994).

To interpret Fig. 2C as the age distribution for life in the Universe
several assumptions need to be made. Among them are:

\begin{enumerate}

\item
Life is based on molecular chemistry and cannot be based on just hydrogen and helium.

\item 
The dominant harbours for life in the Universe are on 
the surfaces of earths in classical habitable zones.

\item
Other time-dependent selection effects which promote or hamper the formation of life  
(supernovae rate?, gamma ray bursts?, cluster environments?) are not as important as the metallicity 
selection effect discussed here (Norris 2000).

\item
Life is long-lived. If life goes extinct on planets then the $PFR$ needs to
have its oldest tail chopped off to represent only existing life.

\end{enumerate}

\section{Discussion}

This paper is an attempt to piece together a consistent
scenario from the most recent observations of
extrasolar planets, the star formation rate of the Universe
and the metallicity evolution of the star-forming regions of the Universe.
The precision of all of these data sets is improving rapidly.
With more than 2000 stars now being surveyed, we expect more than $\sim 100$ giant planets
to be detected in the next few years.
The metallicities of target lists are also under investigation.
Thus, the uncertainties in the metallicities of target stars and stars
hosting planets will be reduced (reducing the error bars in both the
numerator and denominator of Eq. 1).

Planet/planet interactions may explain the hot jupiter/high metallicity
correlation but at least two other (non-mutually exclusive) explanations exist: 
1) metallicity enhanced migration of giants in protoplanetary disks (e.g. Murray \etal 1998) and 
2) infall of metal-rich accretion disks onto the host stars, precipitated by the 
in-spiraling of large planets (e.g. Gonzalez 1998, Quellin and Holman 2000).
The infall of metallicity-enhanced material probably occurs in all 
migration or interaction scenarios. However, if the outer convective 
zones of G dwarfs are thick enough to mix and dilute this material 
(Laughlin and Adams 1999) then the analysis done here requires no 
significant modification for metallicity enhancement. If the dilution 
is not effective then the observed metallicity of a star will not be 
a faithful indicator of the star's true metallicity and therefore will 
not be a good indicator of the probability to produce earths as assumed here.

The results obtained here for the metallicity and age distributions of earth-like planets in the Universe 
are easily testable.
Over the next decade or two, intensive efforts will be focused on finding earths in the solar neighborhood.
Microsecond interferometry (SIM) and even higher angular resolution infrared interferometry (TPF \& IRSI-DARWIN)
as well as  micro-lensing planet searches (PLANET) and high sensitivity transit photometry (COROT) all have the potential 
to detect earth-like planets.
These efforts will eventually yield metalllicity and age distributions for the host stars of earth-like planets
that can be compared to Figs. 1 and 2C.
In addition, Figs. 1 and 2C can be used to focus these efforts.

{\bf Acknowledgemnets}\\
I acknowledge G. Gonzalez, F. Rasio, J. Lissauer,  A. Fernandez-Soto, D. Whittet, 
G. Wetherill,  P. Butler, K. Ragan and R. dePropris
for helpful comments and discussion.
This work has been supported by an Australian Research Council Fellowship.

\begin{center}
References
\end{center}


\hbox{\vtop{\raggedright\hangafter=1\hangindent=0mm\hsize 160mm\strut
Barger, A.J., L.L. Cowie, and E.A. Richards 2000. 
Mapping the Evolution of High-Redshift Dusty Galaxies with Submillimeter Observations of a Radio-selected Sample.
{\it Astrophys. J.} {\bf 119}, 2092-2109. 
\strut}}

\hbox{\vtop{\raggedright\hangafter=1\hangindent=0mm\hsize 160mm\strut
Beckwith, S.V.W., T. Henning, and Y. Nakagawa 2000.
Dust Properties and Assembly of Large Particles in Protoplanetary Disks. 
In {\it Protostars and Planets IV}, (V. Mannings, A.P. Boss and S.S. Russell Eds.)  pp 533-558.
Univ. of Arizona Press, Tucson.
\strut}}

\hbox{\vtop{\raggedright\hangafter=1\hangindent=0mm\hsize 160mm\strut
Bodenheimer, P., O. Hubickyj, and  J.J. Lissauer 2000. 
Models of in Situ formation of Detected Extrasolar Giant Planets.
{\it Icarus} {\bf 143}, 2-14.
 \strut}}

\hbox{\vtop{\raggedright\hangafter=1\hangindent=0mm\hsize 160mm\strut
Boss, A.P. 1995. 
Proximity of Jupiter-like Planets to Low-mass Stars.
{\it Science} {\bf 267}, 360. \strut}}


\hbox{\vtop{\raggedright\hangafter=1\hangindent=0mm\hsize 160mm\strut
Butler, R.P.,S.S. Vogt, G.W. Marcy, D.A. Fischer, G.W. Henry, and K. Apps 2000. 
Planetary Companions to the Metal-Rich Stars BD-10 3166 and HD 52265.
{\it Astrophys. J.} {\bf 545} 504-511 \strut}}

\hbox{\vtop{\raggedright\hangafter=1\hangindent=0mm\hsize 160mm\strut
Favata, F., G. Micela, and S. Sciortino, 1997. 
The [Fe/H] distribution of a volume limited sample of solar-type stars and its
implications for galactic chemical evolution.
{\it Astron. Astrophys.} {\bf 323}, 809-818, Fig. 3 histogram of [$Fe/H$] \strut}}

\hbox{\vtop{\raggedright\hangafter=1\hangindent=0mm\hsize 160mm\strut
Ford, E.B., F.A. Rasio, and  A. Sills 1999. 
Structure and Evolution of Nearby Stars with Planets I. Short-Period Systems.
{\it Astrophys. J.} {\bf 514}, 411-429  \strut}}

\hbox{\vtop{\raggedright\hangafter=1\hangindent=0mm\hsize 160mm\strut 
Gilliland, R.L., and 23 colleagues 2000. 
A Lack of Planets in 47 Tucanae from a Hubble Space Telescope Search.
{\it Astrophys. J.} {\bf 545}, L47-L51 \strut}}  

\hbox{\vtop{\raggedright\hangafter=1\hangindent=0mm\hsize 160mm\strut 
Gonzalez, G. 1997. 
The stellar metallicity -- giant planet connection.
{\it Mon. Not. R. Astron. Soc.} {\bf 285}, 403 \strut}} 

\hbox{\vtop{\raggedright\hangafter=1\hangindent=0mm\hsize 160mm\strut 
Gonzalez, G. 1998. 
Spectroscopic analysis of the parent stars of extrasolar planetary system candidates.
{\it Astron. Astrophys.} {\bf 334}, 221 \strut}}

\hbox{\vtop{\raggedright\hangafter=1\hangindent=0mm\hsize 160mm\strut 
Gonzalez, G. 1999. 
Is the Sun anomalous?
{\it Astron. and Geophys.} {\bf 40} (5), 25 \strut}} 

\hbox{\vtop{\raggedright\hangafter=1\hangindent=0mm\hsize 160mm\strut 
Gonzalez, G. 2000. 
Chemical Abundance Trends Among Stars with Planets.
In {\it Disks, Planetesimals and Planets} ASP Conference Series, Vol. 219,
(F. Garzon, C. Eiroa, D. de Winter, and T.J. Mahoney, Eds.) \strut}} 

\hbox{\vtop{\raggedright\hangafter=1\hangindent=0mm\hsize 160mm\strut
Gummersbach, C.A., A. Kaufer, D.R. Schaefer,  T. Szeifert, and B. Wolf 1998. 
B stars and the chemical evolution of the galactic disk.
{\it Astron. Astrophys.} {\bf 338}, 881-896 
\strut}}
%

\hbox{\vtop{\raggedright\hangafter=1\hangindent=0mm\hsize 160mm\strut
Habing, H.J., C. Dominik, M. Jourdain de Muizon, M.F. Kessler, R.J. Laureijs, K. Leech, 
L. Metcalfe, A. Salama, R. Siebenmorgen, and N. Trams 1999. 
Disappearance of stellar debris disks around main-sequence stars after 400 million years.
{\it Nature} {\bf 401}, 456-458
\strut}}

\hbox{\vtop{\raggedright\hangafter=1\hangindent=0mm\hsize 160mm\strut 
Laird, J.B., Carney, B.W., M.P. Rupen, and D.W. Latham 1988. 
A survey of proper-motion stars. VII The halo metallicity distribution function.
{\it Astron. J.}, 96, 1908 \strut}} 

\hbox{\vtop{\raggedright\hangafter=1\hangindent=0mm\hsize 160mm\strut 
Laughlin, G., and F.C. Adams  1997.
Possible Stellar Metallicity Enhancements from the Accretion of Planets.
{\it Astrophy. J.} {\bf 491} L51-L55 \strut}}  

\hbox{\vtop{\raggedright\hangafter=1\hangindent=0mm\hsize 160mm\strut
Lazcano, A. and S.L. Miller 1994.  
How long did it take for life to begin and evolve to cyanobacteria?
{\it J. Mol. Evol.} {\bf 39}, 549-554 \strut}}

\hbox{\vtop{\raggedright\hangafter=1\hangindent=0mm\hsize 160mm\strut
Lin, D.N.C., P. Bodenheimer, and D.C. Richardson 1996. 
Orbital migration of the planetary companion of 51 Pegasi to its present location.
{\it Nature} {\bf 380}, 606-607 \strut}}

\hbox{\vtop{\raggedright\hangafter=1\hangindent=0mm\hsize 160mm\strut
Lin, D.N.C. and S. Ida 1997. 
On the Origin of Massive Eccentric Planets.
{\it Astrophys. J.} {\bf 477}, 781-791 \strut}}

\hbox{\vtop{\raggedright\hangafter=1\hangindent=0mm\hsize 160mm\strut
Lineweaver, C.H. 1999. 
A Younger Age for the Universe.
{\it Science} {\bf 284}, 1503 \strut}}

\hbox{\vtop{\raggedright\hangafter=1\hangindent=0mm\hsize 160mm\strut
Lissauer, J.J. 1995. 
Urey Prize Lecture: On the Diversity of Plausible Planetary Systems.
{\it Icarus} {\bf 114}, 217-236 \strut}}

\hbox{\vtop{\raggedright\hangafter=1\hangindent=0mm\hsize 160mm\strut
Maher, K.A., and D.J. Stevenson  1988. 
Impact Frustration of the Origin of Life.
{\it Nature} {\bf 331}, 612-614 \strut}}

\hbox{\vtop{\raggedright\hangafter=1\hangindent=0mm\hsize 160mm\strut
Marcy, G. W., and  R.P. Butler   2000. 
Millennium Essay: Planets Orbiting Other Suns.
{\it PASP} {\bf 112}, 137-140 \strut}}

\hbox{\vtop{\raggedright\hangafter=1\hangindent=0mm\hsize 160mm\strut
Mayor, M., and D. Queloz 1995. 
A Jupiter-Mass Companion to a Solar-Type Star.
{\it Nature} {\bf 378}, 355 \strut}}

\hbox{\vtop{\raggedright\hangafter=1\hangindent=0mm\hsize 160mm\strut
Mojzsis, S.J., G. Arrhenius, K.D. McKeegan, T.M. Harrison, A.P. Nutman, and C.R.L. Friend 1996. 
Evidence for life on Earth by 3800 million years ago.
{\it Nature} {\bf 384}, 55-59 \strut}}
 
\hbox{\vtop{\raggedright\hangafter=1\hangindent=0mm\hsize 160mm\strut 
Murray, N., B. Hansen, M. Holman, and S. Tremaine 1998. 
Migrating Planets.
{\it Science} {\bf 279}, 69 \strut}}

\hbox{\vtop{\raggedright\hangafter=1\hangindent=0mm\hsize 160mm\strut
Nishi, R., and M. Tashiro  2000. 
Self Regulation of Star Formation in Low Metallicity Clouds.
{\it Astrophys. J.} {\bf 537}, 50-54, \strut}}

\hbox{\vtop{\raggedright\hangafter=1\hangindent=0mm\hsize 160mm\strut
Norris, R. 2000. 
How old is ET?
In {\it When SETI Succeeds: The impact of high-information Contact} (A. Tough Ed.), Foundation of the Future,
Washington DC
\strut}}

\hbox{\vtop{\raggedright\hangafter=1\hangindent=0mm\hsize 160mm\strut
Papaloizou, J.C.B., and C. Terquem 1999. 
Critical protoplanetary core masses in protoplanetary disks and the formation of short period giant planets.
{\it Astrophys. J.} {\bf 521}, 823-828 \strut}}

\hbox{\vtop{\raggedright\hangafter=1\hangindent=0mm\hsize 160mm\strut
Queloz, D., M. Mayor, L. Weber, A. Blecha, M. Burnet, B. Confino, D. Naef, F. Pepe, N. Santos, and S. Udry  2000. 
The CORALIE survey for southern extra-solar planets I. A planet orbiting the star Gleise 86$^{*}$.
{\it Astron. Astrophys.} {\bf 354}, 99-102 \strut}}

\hbox{\vtop{\raggedright\hangafter=1\hangindent=0mm\hsize 160mm\strut 
Quillen, A.C., and M. Holman  2000. 
Production of star-grazing and star-impacting planetesimals via orbital migration of extrasolar planets.
{\it Astron. J.} {\bf 119} 397-402 \strut}} 

\hbox{\vtop{\raggedright\hangafter=1\hangindent=0mm\hsize 160mm\strut
Rasio, F.A., and E.B. Ford  1996. 
Dynamical instabilities and the formation of extrasolar planetary systems.
{\it Science}  {\bf 274}, 954-956 \strut}}

\hbox{\vtop{\raggedright\hangafter=1\hangindent=0mm\hsize 160mm\strut 
Rocha-Pinto, H.J., and W.J. Maciel  1996. 
The metallicity distribution of G dwarfs in the solar neighborhood.
{\it Mon. Not. R. Astron. Soc.} {\bf 279}, 447-458 \strut}}  

\hbox{\vtop{\raggedright\hangafter=1\hangindent=0mm\hsize 160mm\strut
Schneider, J. 2000. 
Extra-solar planets catalog. 
http://www.obspm.fr/encycl/catalog.html \strut}}

\hbox{\vtop{\raggedright\hangafter=1\hangindent=0mm\hsize 160mm\strut 
Sommer-Larsen, J. 1991. 
On the G-dwarf abundance distribution in the solar cylinder.
{\it Mon. Not. R. Astron. Soc.} {\bf 249}, 368-373 \strut}}

\hbox{\vtop{\raggedright\hangafter=1\hangindent=0mm\hsize 160mm\strut
Trimble, V. 1997. 
Origin of the Biologically Important Elements.
{\it Origins of Life Evol. Biosphere} {\bf 27}, 3 \strut}}  

\hbox{\vtop{\raggedright\hangafter=1\hangindent=0mm\hsize 160mm\strut
Wasserburg, G.J., and Y.-Z. Qian  2000. 
A Model of Metallicity Evolution in the Early Universe.
{\it Astrophys. J.} {\bf 538}, L99-L102 \strut}}

\hbox{\vtop{\raggedright\hangafter=1\hangindent=0mm\hsize 160mm\strut
Weidenschilling, S.J.  1998. 
Growing Jupiter the Hard Way.
American Astronomical Society, DPS meeting \#30, \#21.03 \strut}}

\hbox{\vtop{\raggedright\hangafter=1\hangindent=0mm\hsize 160mm\strut
Weidenschilling, S.J., and  F. Marzari  1996. 
Gravitational scattering as a possible origin for giant planets at small stellar distances.
{\it Nature} {\bf 384}, 619-621 \strut}}

\hbox{\vtop{\raggedright\hangafter=1\hangindent=0mm\hsize 160mm\strut
Wetherill, G.W. 1996. 
The Formation and Habitability of Extra-Solar Planets.
{\it Icarus} {\bf 119}, 219-238 \strut}}

\hbox{\vtop{\raggedright\hangafter=1\hangindent=0mm\hsize 160mm\strut
Whittet, D. 1997. 
Galactic metallicity and the origin of planets (and life).
{\it Astron. and Geophys.} {\bf 38} (5), 8 \strut}}


\end{document}